\documentstyle[aps,prb]{revtex}

\begin{document}
\author{K.D. Belashchenko, V.P. Antropov and S.N. Rashkeev$^\dagger$}
\address{Ames Laboratory, Ames, IA 50011}
\address{$^\dagger$Vanderbilt University, Nashville, TN 37235}
\title{Anisotropy of $p$ states and $^{11}$B nuclear spin-lattice relaxation in
Mg$_{1-x}$Al$_x$B$_2$}
\date{\today}
\maketitle

\begin{abstract}
We calculated the nuclear spin-lattice
relaxation rate in the Mg$_{1-x}$Al$_x$B$_2$ system and found that the
orbital relaxation mechanism dominates over the dipolar and
Fermi-contact mechanisms in MgB$_2$, whereas in AlB$_2$ due to a smaller
density of states and strong anisotropy of boron $p$ orbitals the relaxation
is completely determined by Fermi-contact interaction. The results for
MgB$_2$ are compared with existing experimental data, whereas our results
for the doped alloy and AlB$_2$ can be used in future experiments to
verify the theoretical predictions about the electronic structure of this
system.

\end{abstract}

The discovery of superconductivity in MgB$_2$\cite{akim} has stimulated
a significant interest in the electronic structure of these intercalated boron
compounds. For the superconductivity, one of the most
important parameters is the value $N(E_F)\equiv N$ of the
electronic density of states (DOS) at the Fermi level in the normal state. Several
groups have performed local density-functional (LDA) calculations of this
quantity and the results scatter around 0.7 states/(eV$\cdot$f.u.) \cite{us,LDA}.
Experimentally $N$ can be determined in
many different ways but the nuclear
spin-lattice relaxation (NSLR) rate $1/T_1$ measurements
represent an excellent opportunity to check experimentally not only the total N
and its partial components but also their anisotropy, i.e. the distribution
between the in-plane and out-of-plane $p$ orbitals. Experimental numbers have
been reported in Ref.~\onlinecite{EXP1,EXP2,EXP3} where values of
$TT_1=$ 180, 155 and 165 K$\cdot$sec for MgB$_2$ on $^{11}$B
($\mu=2.689\mu_N$) were obtained. These authors interpreted the relaxation rates
in terms of dipolar and orbital contributions due to the corresponding Korringa
ratio and the already known\cite{us,LDA} theoretical total $N$. Below by
using the general formulas for $T_{1}$ in the hexagonal crystal\cite{OBATA} we
evaluate the relaxation rates for MgB$_{2}$ and AlB$_{2}$
compounds using an LDA calculation. We will show that in MgB$_{2}$
the orbital relaxation rate is about three times larger than the
dipolar and the Fermi-contact rates with all other contribution being much
smaller, whereas the Fermi-contact mechanism is dominating in AlB$_{2}$.
Our calculated total rates are in fair agreement with those measured
\cite{EXP1,EXP2,EXP3}, thus indicating that the calculated $N$ in MgB$_{2}$
is basically correct, whereas in AlB$_{2}$, due to an unusually high anisotropy
of $p$ states, we predict a much smaller relaxation rate fully determined
by Fermi-contact mechanism.

The standard technical details of the LMTO-ASA calculations are similar to those
described in Refs.~\onlinecite{us}. In addition, we carefully checked the
sensitivity of ASA results to the parameters of calculations. We used different
exchange-correlation potentials and inputs with different radii of the
B sphere (both with and without empty spheres). Our values of DOS for $s$ and $p$
orbitals $N_{lm}$ for the B site in MgB$_2$ and AlB$_2$
are listed in Table I. One can see that in MgB$_2$ all $p$ orbitals on the
B site have a sizeable DOS, while in AlB$_2$ only $p_z$ orbital is
significant with $N_{px}\approx0.1N_{pz}$. The $s$ component in AlB$_2$
becomes relatively more important compared to MgB$_2$ resulting in the
dominance of the Fermi-contact NSLR in AlB$_2$, as we will show below.
In both materials the contribution of $d$ states to NSLR is very small.
As for the Mg site, the $s$ component of $N$ is the most important, and
we expect that the NSLR on $^{25}$Mg is controlled by the Fermi-contact
mechanism. However, in this paper we will focus on the $^{11}$B NSLR.

To calculate $T_{1}$ according to a general prescription one has
to estimate the $\langle r^{-3}\rangle_l$ expectation values for different orbitals and
the electronic density at the nucleus $\varphi_s^2(0)/4\pi$. These parameters
(properly normalized with the densities of states) are the largest source of
uncertainty in our calculation due to the notable dependence of
$\langle r^{-3}\rangle_{p}$ on the chosen radius of the B sphere $r_{B}$
(from $r_B=2.15a_0$ to $r_B=2.4a_0$ it decreases by $\sim$11\%).
In our calculations we used the largest $r_B$ that were possible without
a significant distortion of the band structure, 2.4$a_0$ for MgB$_{2}$
and 2.1$a_0$ for AlB$_{2}$. For these radii we have
$\langle(a_0/r)^3\rangle_{p}=1.11$ in MgB$_2$ and 1.37 in AlB$_2$.
Although due to the poor convergence these averages probably still have a
considerable inaccuracy of about 10-15\% (which is close to the experimental
error bar), we believe that they are the most accurate that may be obtained
in ASA. For comparison, the atomic value\cite{Fraga} for
$\langle (a_0/r)^3\rangle_{p}$ in B is 0.78. The electronic densities on
the nucleus $a_0^3\varphi_{s}^{2}(0)/4\pi$ for MgB$_2$ and AlB$_2$ were,
respectively, 2.68 and 3.02. 

The contributions to the $^{11}$B relaxation rate for the polycrystalline
sample calculated using the
formulas of Ref.~\onlinecite{OBATA} are given in Table II.
The in-plane and out-of-plane $p$ orbitals in MgB$_{2}$ have similar
densities of states, and hence the relative magnitude of orbital and dipolar
contributions to NSLR is similar to the 3/10 rule for $p$ states in a
cubic crystal described by Obata\cite{OBATA}. The Fermi-contact
contribution is also important and amounts to 30\% of the orbital term.
The contributions from the $d$ partial waves to the dipole and orbital
relaxation rates were small (at the order of 1\%) due to the low diagonal
and off-diagonal densities of states $(N_d/N_p)^2\sim0.02$ and 
$(N_{pd}/N_p)^2\sim0.05$. The quadrupole contribution to NSLR is negligible
due to a rather small $^{11}$B quadrupole moment.

The values of $T_{1}$ obtained in such manner correspond to the theoretical `bare'
$N$ which does not include the exchange-correlation enhancement.
We have estimated the effective Stoner exchange parameter to be
$I\equiv\Delta E/m=1.7$ eV from the splitting of the bands at the
$\Gamma$ point in the external magnetic field. The corresponding Stoner
enhancement of the uniform susceptibility (which enters the
$T_1^{-1}$) can be written\cite{ENHANS} as $S=3/\left[ (1-NI)\left( 3-2IN\right) \right]
\approx (1-IN)^{-\alpha }$, where $\alpha\approx1.62$ in the 3D case 
and in the 2D case $\alpha=2$. In our case due to the mixed
2D and 3D character of the bands it is not clear what value of $\alpha$
should be used in our simple estimation. We used $\alpha=1.9$, resulting
in the enhancement of $T_1^{-1}$ by approximately 60\% (Table II).
The total calculated relaxation rate 81$\cdot10^{-4}$ K$\cdot$sec for MgB$_2$
should be compared with the experimental rates of 56--64$\cdot10^{-4}$
(K$\cdot$sec)$^{-1}$ measured at temperatures slightly above
$T_{c}$\cite{EXP1,EXP2,EXP3}. The fact that such simple estimate
gives a faster relaxation compared to experiments, even without taking into
account any other mechanisms of susceptibility enhancement, may serve as a
manifestation of a possible importance of unusual effects resulting in
the lowering of the effective $N$.

We obtained a very different relation between the different
mechanisms of NSLR in AlB$_2$ where no traces of superconductivity have
been found and where no NMR data are
available. According to our theoretical estimation, due to the sharp decrease of
the $p$ component of $N$ compared to MgB$_{2}$, the orbital and dipolar
contributions to NSLR become very small, and the total NSLR in AlB$_2$
is completely dominated by the Fermi-contact mechanism. The resulting
NSLR rate is more than two times smaller than in MgB$_2$ (see Table II).
Corresponding
numbers for the anisotropy parameter at the Fermi level $N_{px}/N_{pz}$ are
$0.1$ for AlB$_{2}$ and 0.73 for MgB$_{2}$. This specific feature of AlB$_{2}$
on this stage is completely theoretical prediction and more experimental
information needed to check it. We also performed the rigid band
calculations of $T_{1}$ in Mg$_x$Al$_{1-x}$B$_2$. From Fig.~1
one can see how different mechanisms of NSLR are progressing as a
function of doping. The sharp decrease of $N$ in the 2D sheets of the
Fermi surface\cite{us} leads to a corresponding lowering of all
contributions to NSLR, and at the point of the structural transition
we expect a very large $T_1$. We expect that the new NMR experiments for this 
alloy will be a crucial test of our understanding of the electronic
structure of this system.
Simultaneously with the nuclear quadrupole resonanse data on this material
(which is related to the anisotropy of total charges on different
$p$ orbitals) the general picture of anisotropy of $p$ orbitals in normal states
can be build.

The above calculations have been done for a polycrystalline material.
Because single crystals are becoming available, we include our estimations
of the anisotropy in the angular dependence of NSLR rate\cite{OBATA}
$A+B\sin^2\theta$. For MgB$_2$ we obtained $B/A\approx-0.06$, so that
the NSLR is nearly isotropic.

In conclusion, we have performed LDA calculations for the NSLR rate in the
Mg$_x$Al$_{1-x}$B$_2$ system. We find that the orbital mechanism of relaxation
dominates over the spin-dipolar and Fermi-contact mechanisms in MgB$_2$, because
the boron $p$ orbitals at the Fermi level are distributed nearly isotropically 
and have a large DOS. With the values of 0.7 st./eV for the bare DOS at
the Fermi level for MgB$_2$ and 1.7 eV for the effective Stoner exchange
parameter, reasonable agreement is obtained with the experimental relaxation
rates. We estimate the overall error of our $T_1$ calculations in MgB$_{2}$ as
15\%. Strong anisotropy of $p$ states at the Fermi level and different leading
mechanism of relaxation in AlB$_2$ compared to MgB$_2$ is predicted.  

When this paper was completed we became aware that similar results for NSLR in
MgB$_{2}$ were independently obtained in Ref.~\onlinecite{Mazin}, where the
$T_1$ on $^{25}$Mg and the Knight shifts were also computed, as well as the
small core polarization term for $^{11}$B.

We are grateful to I. Mazin and F. Borsa for valuable discussions.
This work was carried out at the Ames Laboratory, which is operated for the
U.S.Department of Energy by Iowa State University under Contract No.
W-7405-82. This work was supported by the Director for Energy Research,
Office of Basic Energy Sciences of the U.S. Department of Energy.

\begin{table}[tbp]
\caption{Partial DOS for $s$ and $p$ orbitals at B site,
10$^{-3}$ (eV$\cdot$spin$\cdot$atom)$^{-1}$}
\begin{tabular}{l|ccc}
\  & $s$ & $p_z$ & $p_x$\\ 
\tableline MgB$_{2}$ & 3.4 & 50 & 36\\ 
AlB$_{2}$ & 3.3 & 19 & 1.9
\end{tabular}
\end{table}
\begin{table}[tbp]
\caption{Contributions to $(TT_{1})^{-1}$ [$10^{-4}$/(K sec)]}
\begin{tabular}{l|cccccc}
\  & Fermi-contact & Orbital & Dipole & Total & Stoner-enhanced & Experiments\\ 
\tableline MgB$_{2}$ & 12 & 30 & 9 & 51 & 81 & 56\cite{EXP1},64\cite{EXP2},61\cite{EXP3}\\ 
AlB$_{2}$ & 21 & 1 & 1 & 23 & 26 & ---
\end{tabular}
\end{table}

\begin{figure}
\caption{Different contributions to the total $^{11}$B relaxation rate
in Mg$_{1-x}$Al$_x$B$_2$. Thick solid line, orbital; dashed, dipolar; thin solid, Fermi-contact.}
\end{figure}

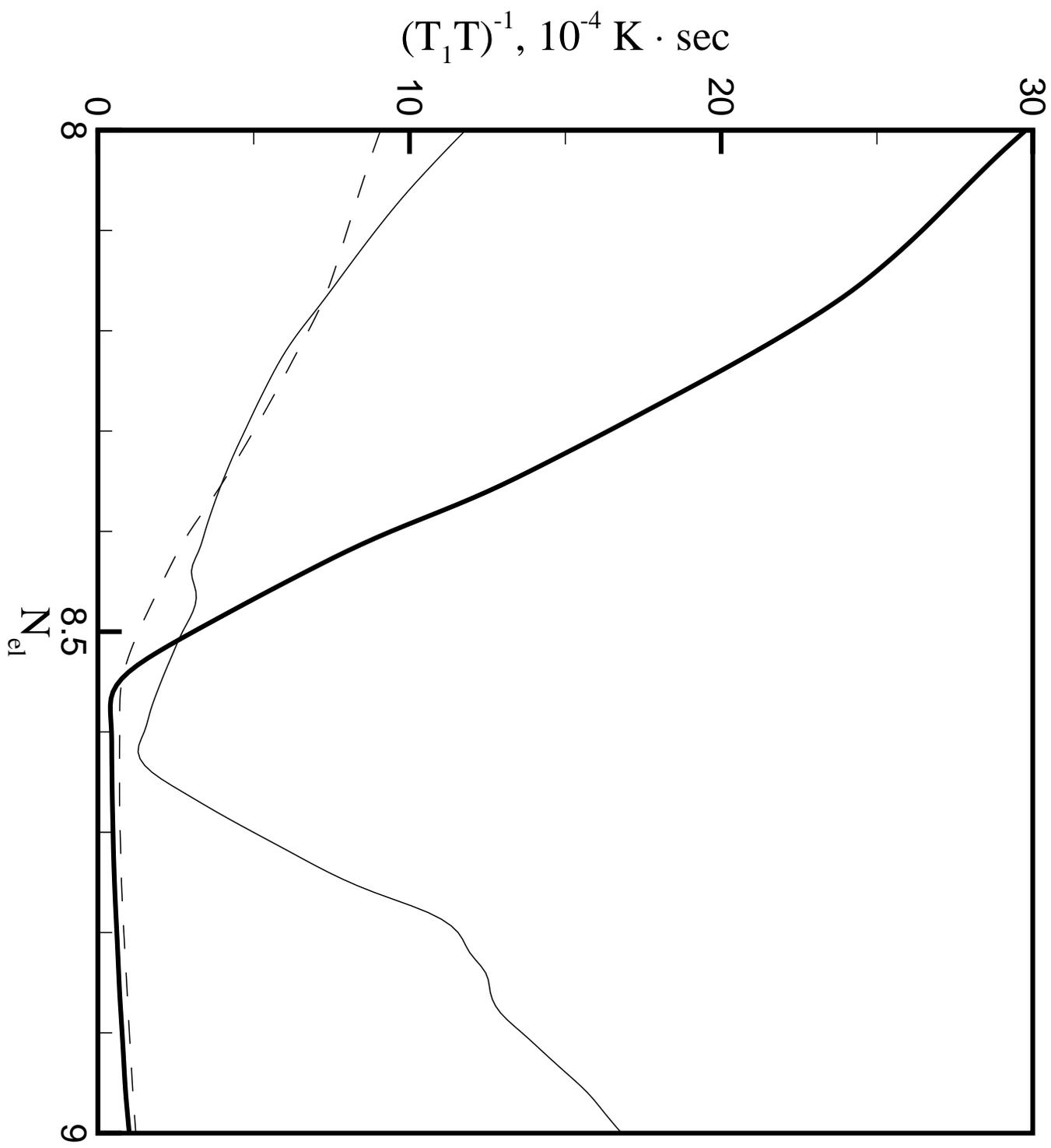
\end{document}